\documentclass[12pt,a4paper]{article}
\usepackage{amsfonts, amsmath, amssymb}
\usepackage{graphicx,xspace,psfrag,color}  
\usepackage{mathrsfs}
\usepackage{geometry,hyperref,latexsym,pifont,textcomp,setspace}
\usepackage[abs]{overpic}

\newcommand{\scri}{\mathscr{I}}

\psfrag{t}[][]{\Large$t$}
\psfrag{r}[][]{\Large$r$}
\psfrag{ip}[][]{$i^+$}
\psfrag{im}[][]{$i^-$}
\psfrag{i0}[][]{$i^0$}
\psfrag{scrip}[][]{$\scri^+$}
\psfrag{scrim}[][]{$\scri^-$}
\title{\Huge\bf  Mutiny at the white-hole district}
\author{
Carlos Barcel\'{o}$^1$,
Ra\'ul Carballo-Rubio$^{1}$,
Luis J. Garay$^{2,3}$\\
\small\it $^1$ Instituto de Astrof\'{i}sica de Andaluc\'{i}a, CSIC,\\
\small\it Camino Bajo de Hu\'{e}tor 50, 18008 Granada, Spain\\
\small\it $^2$  Departamento de F\'{i}sica Te\'{o}rica II,
Universidad Complutense de Madrid,\\
\small\it 28040 Madrid, Spain\\
\small\it $^3$  Instituto de Estructura de la Materia, CSIC,\\
\small\it Serrano 121, 28006 Madrid, Spain
}

\def\fecha{28 March 2014}
\date{\fecha}

\begin{document}
\maketitle
\thispagestyle{empty}


\bigskip
\hrule
\begin{abstract}
\noindent
The white-hole sector of Kruskal's solution is almost never used in physical applications. However, it might contain the solution to many of the problems 
associated with gravitational collapse and evaporation. This essay tries to draw attention to some bouncing geometries that make a democratic use of the black- and white-hole sectors. We will argue that these types of behaviour could be perfectly natural in some approaches to the next physical level beyond classical general relativity.

\end{abstract}
\hrule
\bigskip

\maketitle
 \begin{center}
Essay awarded a honorable mention in the\\ 2014 Gravity Research Foundation essay competition.
 \end{center}
\vfill

\bigskip
{\underline{E-mail}: carlos@iaa.es,~raulc@iaa.es,~luisj.garay@ucm.es}

\clearpage
\markright{Mutiny at the white-hole district \hfil }
\pagestyle{myheadings}

\def\HRULE{{\bigskip\hrule\bigskip}}




 \noindent
 {\bf Introduction:} The prevalent view is that black holes indeed form in astrophysical scenarios and that, by themselves (i.e. forgetting their gravitational effects in surrounding matter),   would remain absolutely inert for very many Hubble times except for a tiny evaporative effect that would eventually make them  disappear. Whether these astrophysical objects are strict general-relativistic black holes (possessing an event horizon) or not has been a matter of controversy for over 30 years~\cite{Hawking1976,Susskind2008} (this is known as the information-loss problem). Nowadays even   Hawking concedes that no strict event horizon would ever form, only long-last trapping horizons~\cite{Hawking2014}. Whereas the question is of fundamental interest on purely theoretical grounds, it is almost certainly irrelevant for all astrophysical purposes (nature might still surprise us with the existence of primordial black holes).
 
 Different evaporative scenarios have been designed to accommodate solutions to the information problem (see e.g. Susskind~\cite{Susskind1993,Susskind2012}, Mathur~\cite{Mathur2005}, Dvali~\cite{Dvali2012}, Giddings~\cite{Giddings2012}). More recently, inspired by loop quantum cosmology ideas, Rovelli and Vidotto have put forward an evaporation model such that when the collapsing matter approaches the classical would-be singularity it undergoes a kind of quantum bounce~\cite{Rovelli2014} (there is a similar proposal by Bambi et al~\cite{Bambi2014} and an
early one by Ashtekar and Bojowald~\cite{Ashtekar2005}). In all  these scenarios  the astrophysical view of slowly evaporating trapping horizons is untouched.  
 
We want to draw attention to a completely different take on the issue. Our proposal goes further than that of Rovelli-Vidotto by regularizing the classical would-be singularity by an absolutely genuine bounce. It requires one assumption about the physics of the next layer beyond general relativity, promted when Planck densities are approached, and that independently of its specific structure.
Before making the actual proposal let us recall the current situation in the white-hole district.       
 
 \vspace{0.5cm}
 \noindent
 {\bf The white-hole district:} The simplest models of collapse are   spherically symmetric and we will concentrate on them.
 Birkhoff's theorem is seldom thought of as a  uniqueness result. This is certainly the case for the vacuum geometry outside a spherical distribution of matter   exterior to the Schwarzschild radius. However, for vacuum geometries inside the Schwarzschild radius this is not true. Birkhoff's theorem asserts that any vacuum patch must be locally equivalent to a patch of the maximally extended Kruskal solution~\cite{Hawking1973}. The vacuum geometry outside a spherical distribution of matter but still inside the Schwarzschild radius can be of black-hole or of white-hole type. The Schwarzschild interior regions are dynamical and can be either contracting towards a singularity (black-hole district) or expanding from a singularity (white-hole district). Birkhoff's theorem does not tell us which internal geometry is the appropriate one.
 
Within general relativity stellar objects eventually collapse to form black holes and future singularities. This is arguably why the white-hole region is considered as non-physical and is almost forgotten in what, no surprise, happens to be known as ``black hole physics''.
 
What would  happen if the collapsing matter undergoes a perfectly time-sym\-me\-tric genuine bounce when reaching Planck densities? For example, within loop quantum gravity it is not difficult to imagine the existence of an effective Newtonian potential with a repulsive core (e.g.~\cite{Rovelli2014}) 
 \begin{eqnarray}
 	V_{\rm eff}= -\frac{m}{r} + \frac{\lambda}{r^4}~.
 \end{eqnarray}
In Newtonian physics, this kind of potentials will lead to genuine bouncing solutions. General-relativistic situations in which a lump of matter is climbing a gravitational potential can easily be found but, although white-hole solutions exist, never when this climbing occurs beyond the Schwarzschild radius.
The fading into oblivion of the white-hole district is arguably the strongest departure of general relativity from Newtonian physics. There seems to be  a prejudice against exploring the possibility and implications of really having genuine bouncing solutions. We will see that there are indeed good reasons for this prejudice to exist, but also for taking the bold leap of exploring beyond it.
 
\begin{figure}[H!]%
\vbox{ \hfil \includegraphics[width=.9\textwidth]{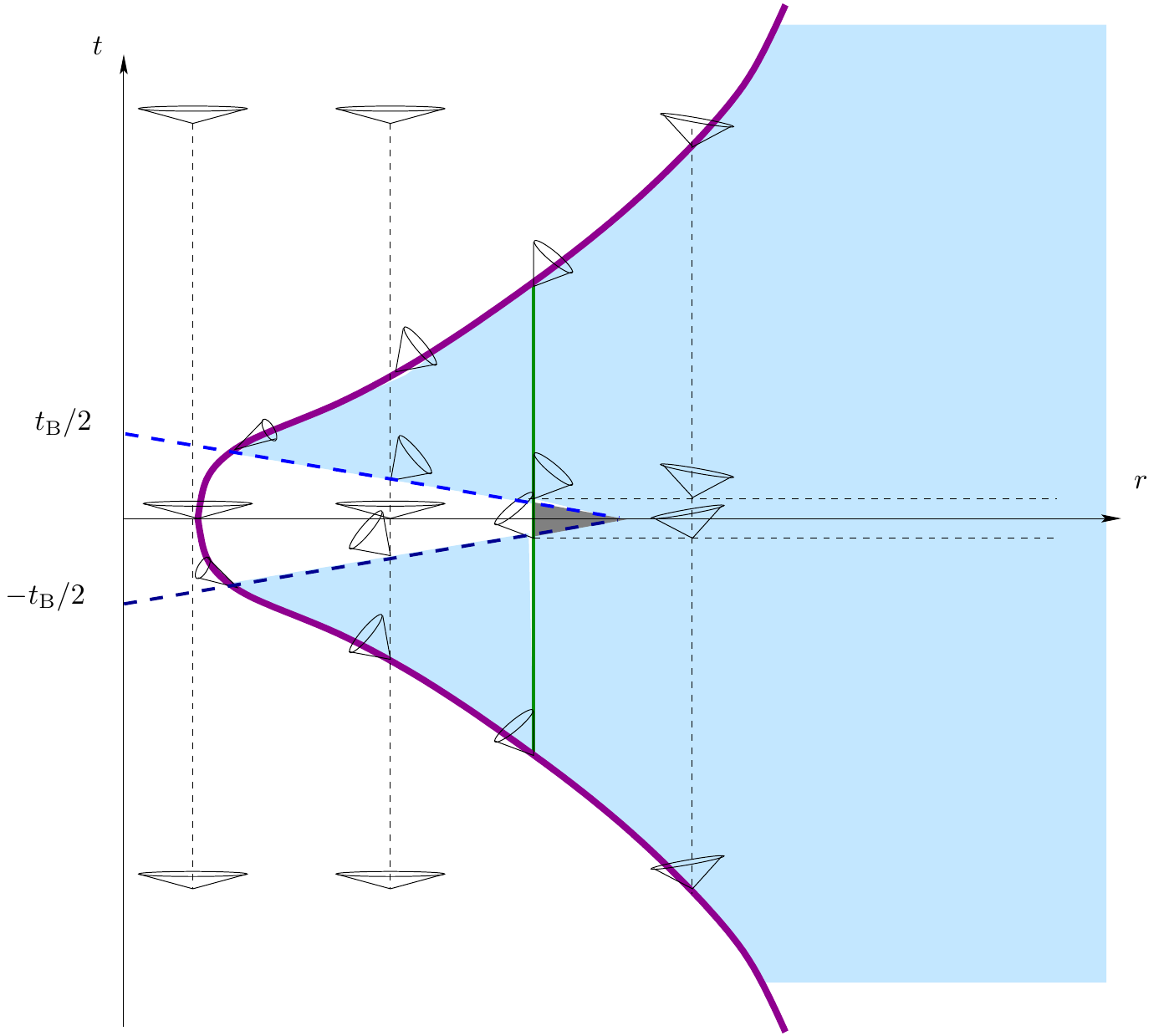}\hfil}%
\bigskip%
\caption{The figure represents the collapse and time-symmetric bounce of an stellar object in our proposal (the thick magenta line). The past dashed blue line marks the boundary where the non-standard gravitational effects start to happen. In all the light-blue region the metric is Schwarzschild. The region between the two dashed blue lines outside matter the metric is not Schwarzschild, including the small grey triangle outside the Schwarzschild radius (the vertical green straight line). The drawing tries to capture the general features of any interpolating geometry.}
\label{Fig:bh-wh-diagram}%
\end{figure}

 \vspace{0.5cm}
 \noindent
 {\bf The proposal:} We propose that when the collapsing matter reaches Planck density it slows down and bounces back connecting with the time-reversed geometry associated with a white-hole spacetime.  Figure~\ref{Fig:bh-wh-diagram} represents such geometry for a homogeneous dust star (Oppenheimer-Snyder model~\cite{Oppenheimer:1939ue}) collapsing from far above its Schwarzschild radius $r_{\rm  {S}}$. The   region $t<-t_{\rm  {B}}/2$ can be described by an acoustic metric
 \begin{eqnarray}
 	ds^2=-(c^2-v^2)dt^2- 2v dtdr + dr^2+r^2 d\Omega_2^2,
 	\label{acoustic-metric}
 \end{eqnarray}
 with
 \begin{eqnarray}
 	c=1,\qquad v =v_{\rm  {I}}:= \left\{
 	\begin{array}{l}
 		-\sqrt{\frac{2M}{ r}},~~~ r>r_s(t) ,
 		\\
 		-\frac{r}{r_0} \left(1-\frac{t}{t_0}\right)^{-1},~~~ 0<r<r_s(t), 
 	\end{array}
 	\right.
 	\label{velocity-profile}
 \end{eqnarray}
and $r_s(t)$ representing the trajectory of the surface of the star. For $t>t_{\rm  {B}}/2$ the metric has this same form  with    $v=v_{\rm  {E}}:=-v_{\rm  {I}}$. The geometry during the short interval $t_{\rm  {B}}$ is  smooth and time-symmetric, and  interpolates the future and past patches.  $t=0$ represents the hyperplane of time-reversal symmetry. 
 
In the wedge between the two dashed blue lines, the light-cone structure smoothly turns towards its time-reversed version. This can be encoded in an acoustic metric with an appropriately symmetrized function $v(t,r)=-v(-t,r)$ and smooth matchings at the dashed blue lines. When these lines reach the Schwarzschild radius (vertical green straight line), one can prescribe acoustic profiles $v(t,r)$ and $c(t,r) \neq 1$ for the area around $t=0$ such that only in the small grey triangle the geometry is actually different from Schwarzschild's. Indeed, the change of coordinates    
 \begin{eqnarray}
 	t'=t + \int_{r_{\rm  {R}}}^r \frac{2v(t,r)}{1-v^2(t,r)} dr,\qquad r'=r,
 \end{eqnarray}
 with $r_{\rm  {R}}>r_{\rm {S}}$ being an arbitrary reference radius and $v(t,r)$ a function interpolating in time between $v_{\rm  {I}}$ and $v_{\rm  {E}}$, changes the metric (\ref{acoustic-metric}) from $v=v_{\rm {I}}$ to   $v=v_{\rm  {E}}$ without changing the geometry. These changes of coordinates would be ill-defined at the horizon itself. Therefore, there must be always a region enclosing a portion of the Schwarzschild radius where the geometry is not Schwarzschild. Physically this is easy to understand. Reversing the light-cone tilting at the Schwarzschild radius is a physical process that mutates an initially marginally trapped surface to a non-trapping surface. However, outside the Schwarzschild radius there exist reversals of the light-cone bending that are just coordinate artifacts. 
 
 \vspace{0.5cm}
 \noindent
 {\bf The underlying hypothesis:} The past dashed blue line marks the boundary where non-standard gravitational effects start to happen. It is born at $r=0$ and travels outwards even though the light cones are pointing inwards. This signal, should it exist, cannot follow the causality associated with the gravitational light cones. Rather it must follow an underlying causality that is explored only when Planck energies are at stake. Otherwise  general-relativistic light cones could not suffer such a dramatic turn: the only possibility left would be that the light cones did not quickly reverse its tilting but only slowly recover their unbent positions before collapse. This is precisely Rovelli-Vidotto's evaporation model. They consider the slow growing (as seen from outside) or anti-evaporation of the internal horizon as a kind of quantum bounce. 
 
Instead, our proposal  connects naturally with a formulation of general relativity as a non-linear graviton theory in a flat Minkowski background (see ~\cite{Carballo2014} and references therein). It is still an open possibility that an ultraviolet completion of such a theory would exhibit asymptotic freedom (as its QCD cousin). In the long-wavelength limit general relativity is recovered, with the general-relativity metric controlling the effective causality in the system. When high-energy phenomena are involved it is reasonable to expect that the underlying causality in the system, which is Minkowskian with no horizons whatsoever, would be unveiled. Similar ideas also appear when thinking on gravity as an emergent notion~\cite{Barcelo2010}. There, it is reasonable to think that at some Planckian energy scale gravity is switched off, still leaving a Minkowskian background for the matter excitations.
 
Thus, for our proposal to make sense we need the assumption that the causality encoded in the general-relativity metric is just a low-energy notion and that at higher energies there is a deeper causality, which from the low-energy perspective would be seen as a superluminar behavior. This superluminarity needs not confront with special relativity, quite the contrary, it could find a robust support on it.

 \vspace{0.5cm}
 \noindent
 {\bf Observational effects:}
 If our idea is at work in nature it will have many observable effects. In the collapse of, say, a neutron star, matter would remain apparently frozen at the Schwarzschild radius for just a few milliseconds before being expelled again. In realistic situations the bounce will be not completely time-symmetric: part of the matter will go through towards infinity while another part would tend to collapse again. The core mass would search for a new equilibrium stage. We do not know the nature of this new metastable state. Most probably it will be small, dark, and heavy, but without black- or white-hole districts~\cite{Visser2009}.

 \newpage

\pagebreak

\end{document}